\documentstyle[12pt]{article}
\begin{document}

{Supermassive black holes may be limited by the Holographic Bound}

\bigskip

\centerline{Paulo Sergio Custodio and J.E. Horvath}

\centerline{Instituto de Astronomia, Geof\'\i sica  e Ci\^encias Atmosf\'ericas, 
Universidade de S\~ao Paulo} 
\centerline{R. do Mat\~ao 1226, Cidade Universit\'aria, 05508-900 
S\~ao Paulo, Brazil}

\bigskip
{Supermassive Black Holes are the most entropic objects
found in the universe. The Holographic Bound (HB) to the entropy is used to
constrain their formation time with initial masses
$\sim{10}^{6-8}M_{\odot}$, as inferred from observations. We
find that the entropy considerations are more limiting than
causality for this "direct" formation. Later we analyze the
possibility of SMBHs growing from seed black holes. The growth of
the initial mass is studied in the case of accretion of pure
radiation and quintessence fields, and we find that there is a
class of models that may allow this metamorphosis. Our analysis
generalizes recent work for some models of quintessence capable of
producing a substantial growth in a short time, while
simultaneously obeying the causal and Holographic Bound limits.}

\bigskip
{\bf Supermassive Black Holes in the Universe}

The continued observations of galaxies has revealed a hidden
population of huge massive objects in compact nuclear regions of
size $\leq$ few pc. Dynamical measurements taken along the last
decade point out that the compact objects have masses in the range
$10^{6}-10^{8} \, M_{\odot}$, and perhaps more importantly, that
every galaxy seems to host a central massive object [1]. Even
though some exotic alternatives have been proposed for their
nature (e.g. neutrinoballs, see [2]), the simplest explanation
is that the central parsecs of the galaxies are sites of residence
of supermassive black holes (SMBH). Among the possible formation
scenarios a hierarchical merging of smaller black holes has been
suggested [3], although it is not guaranteed that the efficiency
of the merging process is high enough to provide large masses. A
likely alternative is that the SMBHs are primordial, i.e. preexist
the galaxies, and perhaps are important for their very formation [4].

While the mass budget of the universe is not likely to be affected
by the presence of the nuclear SMBHs, the total entropy will
certainly be, since the entropy content of the black holes is huge.
This suggests a connection between the formation of the SMBHs and
the total entropy, possibly limited by the Holographic Bound, which has
been proposed to limit the entropy enclosed in a given volume 
and may be deeply related to the fundamental theories [5].

We discuss in this work the issue of entropic limitations to the
formation of SMBHs with masses $\geq 10^{6} \, M_{\odot}$.
After a brief presentation of the Holographic Bound and related
concepts in Section 2, direct formation of SMBHs is addressed in Section 3. 
We analyze the possibility of growing the "seed" black holes to 
those large values is addressed in Section 4. Conditions for fast growth due
to accretion of a quintessence scalar field are addressed in
Section 5.  Section 6 discusses the role of causality in the process of 
accretion. Some general conclusions are given in Section 7.

\bigskip
{\bf The Holographic Bound}

The Holographic Bound may be formulated by asserting that for a
given volume $V$, the state of maximal entropy is the one
containing the largest black hole that fits inside $V$, and this
maximum is given by the finite area that encloses this volume.
This idea generalizes a conjecture made by Bekenstein [6] in
which this maximum is fixed by the non-gravitational energy within
a sphere of size $R$, i.e. $S < {2 \pi E R \over{hc}}$
(now being properly called the {\it Bekenstein limit}).

Several analysis made in recent years reformulated this conjecture
and proposed slightly different forms for the HB, but rather than discussing which
one is correct we will base our argument on the very existence
of some entropy bound, yet to be definitively identified.

To be concrete we shall assume the entropy $S$
to be bounded by the Bekenstein-Hawking value

\begin{equation}
S\leq {A\over{4}}
\end{equation}

where $A$ is the area of the enclosed system under consideration.
Unless explicitly indicated, we shall use natural units throughout
this paper, then the Planck length $L_{planck}^{2} = 1$ in eq.(1)
above and so on.

Verlinde [7] observed some time ago that this bound must be modified in a
cosmology with an arbitrary number of dimensions. Considering the
Einstein space-time with the metric

\begin{equation}
{ds}^{2}=-{dt}^{2}+{R}^{2}{d\Omega_{n}}^{2}
\end{equation}

where ${d\Omega_{n}}^{2}$ is the line element of a unit n-dimensional
sphere, the entropy of the conformal field in this space-time can be
expressed in terms of its total energy $E$ and the Casimir $E_{C}$ by a
generalized form of the Cardy-Verlinde formula as

\begin{equation}
S={2 \pi \over{n}}R \sqrt{E_{C}(2E-E_{C})}
\end{equation}

For a $(n+1)$ dimensional closed universe, the FRW equations are

\begin{equation}
{H}^{2}={16\pi{G}_{n}\over{n(n-1)}}{E\over{V}}-{1\over{R^{2}}}
\end{equation}

\begin{equation}
\dot{H}=-{8\pi{G_{n}}\over{(n-1)}}\biggl({E\over{V}}+P\biggr)+{1\over{R^{2}}}
\end{equation}

where $H(t)={\dot{a}\over{a}}$ is the Hubble parameter (describing the
expansion/contraction of the universe), the dot stands for
differentiation with respect to the proper time, $E$ is the total energy
of matter filling the universe, and $G_{n}$ is the Newton constant in
$(n+1)$ dimensions. $a(t)$ describes the scale factor of the Universe and
$R(t)\propto{a(t)}$ its physical size.

The FRW equation can then be related to three cosmological
entropy bounds; the Bekenstein-Verlinde bound
$S_{BV}={2\pi\over{n}} E R$, the Bekenstein-Hawking bound
$S_{BH}=(n-1){V\over{4G_{n}R}}$ (expressing that the black hole
entropy is bounded by the area of the cosmological model), and the
Hubble bound $S_{H}=(n-1){H V \over{4G_{n}}}$ (a reflection of the
fact that the maximal entropy is produced by black holes of the
size of Hubble horizon). At a critical point defined by $H R = 1$,
all these three entropy bounds coincide with each other. Let us
define $E_{BH}$ such that
$S_{BH}={(n-1)V\over{4G_{n}R}}={2\pi\over{n}}E_{BH}R$. Then, the
first FRW equation takes the form

\begin{equation}
S_{H}={2\pi{R}\over{n}}\sqrt{E_{BH}(2E-E_{BH})}
\end{equation}

which is precisely of the same form as the Cardy-Verlinde formula. Its maximum reproduces the Hubble bound

\begin{equation}
S_{H}\leq{2\pi{R}\over{n}}E
\end{equation}

Therefore, in some sense it may be said that the FRW dynamics "knows" the value of the
maximum entropy filling the universe. This connection between
geometry and dynamics is a consequence of the Holographic
Principle. However, and suggestive as these arguments are, we do not intend
to further analyze them. It is just enough to keep in mind that independently of its specific form, an Holographic Bound is likely to hold for the actual 3+1 universe.

Another important concept needed for the study of SMBHs
is the generalized second law of thermodynamics, formulated by
J.Bekenstein [8] using a series of gedanken experiments. The
generalized second law attempted to cure serious problems
with the matter + radiation entropy as the latter was absorbed onto black
holes (thus causing a growth of the black hole mass).
Given that black holes need a few macroscopic parameters (mass,
angular momentum and charge) for their description, the absorption
of matter+radiation seemed to lead to a decrease of the entropy of
the universe, since the matter+radiation entropy ended hidden
behind the horizon. This was very problematic, since that this
kind of {\it Geroch} process seems to go against the second law of
thermodynamics $\Delta{S} > 0$. Bekenstein conjectured that the
total entropy of the universe plus $N$ black holes is given by the
sum of the matter+radiation entropy, plus the black hole entropy (which is
proportional to the horizon area) in what is now known as the generalized
second law (GSL). The GSL takes the form

\begin{equation}
S_{total}=S_{m+r}+{1\over{4}}\sum_{i}^{N}A_{i}
\end{equation}

where the first contribution is the entropy associated to
usual matter and radiation, and the second term describes the
black hole contribution to the total entropy. Note that the
entropy of just one black hole is numerically huge,
$S_{bh}\sim{10}^{77}{(M/M_{\odot})}^{2}$, and this fact will be
very important to set astrophysical constraints. According to the
GSL, as long as we deal with classical process involving black
holes and matter, the total variation of entropy must be positive

\begin{equation}
\Delta{S_{total}}>0
\end{equation}

In the next Sections we will evaluate some constraints to the mass and formation
time of supermassive black holes using the concepts of HB and the GSL.

\bigskip
{\bf SMBH and Primordial Black Holes: direct formation}

While the exact origin of the SMBHs is not known, it is possible that either a primordial
process contributed to form them as they are, and that they have grown explosively
from a seed population.

Let us discuss direct formation first. It is well-known that big black holes
have a huge entropy, and if the bounds to the cosmic entropy apply, restrictions would arise for a "direct formation"mechanism.
Actually, if we impose

\begin{equation}
S_{smbh}(t_{i})\sim{2\times{10}^{77}}N{(M/M_{\odot})}^{2} <
S_{H}(t_{i})
\end{equation}

where $t_{i}$ stands for the formation time hereafter, and it is further assumed that all black holes form more or less simultaneously,
we may find an upper bound to the SMBH directly formed allowed by the HB. We start by evaluating $S_{H}(t)$ from eq.(7) for a 3+1  FRW universe. We identify
$R$ with the particle horizon $R=R_{hp}(t)\propto{t}$, and the total energy contained within this radius
$E={4\pi\over{3}}\varrho(t)R^{3}_{hp}(t)$. We further restrict the
analysis to the epochs in which $a(t)\propto{t}^{n}$.
Inserting all the quantities, we obtain
$E_{hp}(t)\propto{a^{-4}(t)t^{3}}\propto{t}$, for $n=1/2$. We do
not need consider other forms of entropy in eq.(10), since
black holes actually dominate the entropy budget by a large factor.

Then, multiplying by the particle horizon, we express the HB in
the radiation-dominated era as $S_{hp}(t) \sim {S_{hp}(t_{D}){(t/t_{D})}}^{2}$. For
the matter-dominated era $n=2/3$ an analogous procedure
yields $E_{hp}(t) \propto {\varrho_{m}(t)R^{3}_{hp}(t)}
\propto{t}$. Therefore, the entropy contents become
$S_{hp}(t)\sim{S_{hp}(t_{0}){(t/t_{0})}^{2}}$, with $t_{0}\sim{H_{0}}^{-1}$ the present age of the universe ($\sim{10}^{17}{h_{0}}^{-1}s$) and $t_{D}$ ($\sim{10}^{13}s$) is the radiation-matter decoupling time. For the sake of generality the dimensionless Hubble constant  $h_{0}=(H_{0}/100km{s}^{-1}{Mpc}^{-1})$ has not been fixed, although recent measurements suggest $h_{0}\sim{0.65}$.
We also know that $S_{hp}(t_{0})\sim {8\times{10}^{121}}$, therefore $S_{hp}(t_{D})\sim{8\times{10}^{121}\over{{(t_{0}/t_{D})}^{2}}}$.
Then, for the radiation-dominated era the condition
$S_{smbh}(N,M)< S_{hp}(t)$ yields the maximum SMBH mass allowed by the entropy bound

\begin{equation}
M_{smbh}(t_{i}) < 5.6 \times {10}^{-3} h_{0}(t_{i}/s){({10}^{11}/N)}^{1/2} \, M_{\odot}
\end{equation}

if their number $N$ is equal to the number of galaxies and all them have been assumed to be of the same mass. The SMBHs can form directly only after

\begin{equation}
t_{min}(M) > 5 \times{10}^{7}{(M/{10}^{8}M_{\odot})} {(N/{10}^{11})}^{1/2} \, s
\end{equation}

because before $t_{min}$ the entropy of these $N$ black holes would be
larger than the entropy allowed by the Holographic Bound.

The same reasoning as above can be applied to the formation in the matter-dominated era with the result

\begin{equation}
M_{smbh}(t_{i}) < 0.6 h_{0}^{2}(t_{i}/s){({10}^{11}/N)}^{1/2} \, M_{\odot}
\end{equation}

and an earliest formation time

\begin{equation}
t_{min}(M)>1.8\times{10}^{9}h_{0}^{-1}{(M/{10}^{8}M_{\odot})} {(N/{10}^{11})}^{1/2}  \, s
\end{equation}

Since the mass inside the horizon in the radiation-dominated era is just $M_{hor}(t)\sim {7.6 \times {10}^{37}}(t/1s)\, g \,\, $, and achieves $\sim 10^{15} \, M_{\odot}$ at its very end, we conclude that the availability of entropy is {\it more} restrictive than the demand of a causal formation. In other words, it is not sufficient to have a large horizon in which the SMBH can fit, to be allowed by the entropy of the HB seems to be even more important than that primary requirement.

A realistic and complete model would take into account an Initial Mass Function (IMF) for these black holes, with a general form given by $({dN\over{dM}})$. In this case we would need to replace the formula above by an integral of the form $S_{smbh}(N,M)\sim 2\times{10}^{77}\int_{M_{1}}^{M_{2}}dM({dN\over{dM}}){(M/M_{\odot})}^{2}$, with $M_1$, $M_2$ the lower and upper limits to the masses determined by the specific physical conditions at formation. We shall not address this complex model here and leave it for future work.  We have just considered a delta-type $({dN\over{dM}})\propto{\delta(M-M_{*})}$,
leading to the simple model given by $S_{smbh}(N,M)\propto{N{(M/M_{\odot})}^{2}}$ in this paper.

\bigskip
{\bf SMBH and Primordial Black Holes: grow of seeds to the SMBH scale}

If the direct formation of SMBHs is difficult, one may wonder if there is still a possibility of starting with black holes of masses $M_{i} \ll 10^{6} \, M_{\odot}$ which grow subsequently by accretion. Considering the absorption-evaporation processes of PBHs, we can
identify epochs in which these objects grow or evaporate. The complete evolution of PBH mass is given semiclassically by

\begin{equation}
\biggl({dM\over{dt}}\biggr)=-{D\over{M^{2}}}+B M^{2}c[\sum_{i}\varrho_{i}(t)]
\end{equation}

where the term $c [\sum_{i}\varrho_{i}(t)]$ describes the flux of
whatever component flows through the black hole horizon, $D \sim
{10}^{26}g^{3}s^{-1}$ is the evaporation constant (which, strictly
speaking, depends somewhat on the number of degrees of freedom of
the incoming material) and
$B={27\pi{G}^{2}\over{c^{4}}}\sim{4.6\times{10}^{-55}{cm}^{2}g^{-2}}$
is the absorption constant (related to the cross-section of the
black hole).

When writing down the semiclassical eq.(15) we have not ruled out
any "fuel" contributing to the growth of the black hole
(quintessence has been proposed by Bean and Magueijo in [9]),
provided their flux is large enough to contribute to the mass
balance. If the black holes formed in the radiation-dominated or
matter-dominated era, the main contribution to the second term is
the flux of background radiation (quintessence will be explicitly
addressed later). If we consider the radiation only, the balance
of the r.h.s. terms define the {\it critical mass}, (see Refs.[11-
13]) for the instantaneous equilibrium between black holes and
radiation, and its value is

\begin{equation}
M_{c}(t) \sim {10^{26} \, g \over{T_{rad}(t)/T_{0}}}
\end{equation}

where the radiation temperature $T_{rad}(t)\propto{a(t)^{-1}}$ falls along the cosmological
expansion and $T_{0}$ is the present temperature of the CMBR. If, say,  $t_{i} \sim {1s}$,
the critical mass is then very small, and all the PBHs
candidates to grow to SMBH were well above this instantaneous equilibrium mass value.
Therefore, the Hawking radiation was negligible for them [12].
If the PBHs are feed, they may grow with time,
and the question is whether they can gain mass until the supermassive regime $M \geq {10}^{6}M_{\odot}$ is reached.

It is generally agreed that the gas and dust accretion are not
likely to be important at early times (and do not have simple
behavior with the cosmological time either). However, the
radiation is always absorbed and might be important if the
accretion rate is high enough [12,13]. Neglecting the gas and dust
fuels, and keeping only the radiation we may obtain analytic lower
limits to the growth.

A first specific question we want to address is the following : if one black hole formed
initially at $t_{i}$ satisfies the HB, i.e.
$S(M(t_{i}))\sim{10}^{77}{M(t_{i})}^{2} < S_{hp}(t_{i})\sim{8\times{10}^{121}{(t_{i}/t_{0})}^{2}}$,
is it automatically guaranteed that this object will satisfy the HB all the time?

To answer this question let us consider the most general flux $F(\varrho(t))$ absorbed by
the event horizon by this black hole. Then, the mass accretion rate is given by

\begin{equation}
\biggl({dM\over{dt}}\biggr)={27\pi\over{4}}{r_{g}}^{2}F(\varrho(t))
\end{equation}

If we choose $\varrho=\varrho_{rad}$, then $F=c\varrho_{rad}(t)$; for quintessence accretion we use $F(\varphi)={{\dot{\varphi}}^{2}\over{2}}$, see [9].

Solving formally the eq.(17) above, yields

\begin{equation}
M(t)={M_{i}\over{[1-{27{\pi}M_{i}\over{M^{4}_{pl}}}
\int_{t_{i}}^{t}dt^{\prime}F(\varrho(t^{\prime}))]}}
\end{equation}

On the other hand, these black holes evolve obeying the HB if the
local flux satisfies

\begin{equation}
\int_{t_{i}}^{t}dt^{\prime}F(\varrho(t^{\prime}))<
{{M_{pl}}^{4}\over{27\pi{M_{i}}}}
\biggl[{1-{(M_{i}/M_{\odot}){(t_{0}/t)}\over{2\times{10}^{22}}}}\biggr]
\end{equation}

For the radiation flux, $F(t)=c \varrho_{rad}(t)$ and therefore
eq.(19) requires

\begin{equation}
\varrho_{rad}(t_{i}) < 6.3\times{10}^{28}g{cm}^{-3}{({10}^{15}g/M_{i})}{(t_{i}/s)}
\end{equation}

For $t_{i} \sim {1s}$ and $M_{i} < 1 M_{\odot}$, typical of the
radiation-dominated era, this condition is indeed satisfied.

 The above general results suggest that the global constraint $S_{total}
< S_{hp} \propto {t}^{2}$, (i.e, the HB) implies some kind of
restriction to the cross-section for the {\it local} accretion
onto the black hole, as if the total flux through the horizon
event would have to be modified. A detailed study of the flow into
the black hole is needed to address this issue.

\bigskip
{\bf Quintessence models and the SMBH growth}

Recent work by R.Bean and J.Magueijo [9] suggested an important
growth of seed PBHs when a quintessence scalar field $\varphi$
dominates the accretion. The key new ingredient is the
role played by the kinetic term ${{{\dot{\varphi}}^{2}\over{2}}}$
in the flux onto the PBHs, which is absent in the case of
absorption of pure radiation. We shall describe some growth
solutions of the Bean-Magueijo model of quintessence around SMBHs.
The evolution of the mass of these objects is given by the
following formula

\begin{equation}
\biggl({dM\over{dt}}\biggr)={\eta}CM^{2}{\dot{\varphi}^{2}}
\end{equation}

where $C={27\pi\over{2{M_{pl}}^{4}}}$ and $\eta \leq 1$ is a
parameter that measures the efficiency of the accretion process.
Because of the uncertainties on the value of $\eta$ (related to
the details of the depletion of material nearby the black hole,
see Ref.[10] for a recent discussion), we have left it free in our
calculations, so that the adoption of a different value can be
easily done. The connection between the potential $V(\varphi)$ and
$\dot{\varphi}$ comes because the latter regulates the expansion
rate and hence the behavior of the flux ${\dot{\varphi}^{2}}$. The
complete system to solve is given by eq.(21) above together with
the dynamical equations

\begin{equation}
\ddot{\varphi}+3H\dot{\varphi}+V^{\prime}=0
\end{equation}

and

\begin{equation}
H^{2}={8\pi\over{3{M_{pl}^{2}}}}[V(\varphi)+\varrho_{pbh}]
\end{equation}

This set of differential equations is very difficult to solve,
even in the approximation $\varrho_{pbh}<< V(\varphi)$, which is
relevant for this work. Bean and Magueijo analyzed one
particular model of quintessence in which
$V(\varphi)=\lambda{\exp[\lambda{\varphi}]}$, implying
$\dot{\varphi}^{2}\propto{t}^{-2}$. Within this model, in which
the quintessence flux around the SMBH decreases with time, they
claim that the black holes grow in time. However, the decreasing
quintessence flux around these objects casts doubts on this
result, since the mass gain term decreases accordingly.
We pointed out elsewhere that in the radiation-dominated era the 
growth of black holes is actually quenched when the background flux 
decreases with time as ${t}^{-2}$ or faster (see Ref.[12]).

The question is whether there are PBHs that gain substantial mass
at asymptotic times for a given potential $V(\varphi)$ which determines
the quintessence flux. Let us show a class of solutions involving
quintessence accretion only which can make the PBHs grow, as an example of this general behavior.
The quintessence models satisfying

\begin{equation}
\dot{\varphi}^{2}=M_{pl}^{4}{(t/t_{*})}^{n}
\end{equation}

constitute a class of interesting growing models (we used natural units
and $t_{*}={E_{*}}^{-1}$ is a time constant). For this choice,
the kinetic energy ${{\dot{\varphi}}^{2}\over{2}}= K {t}^{n}$ with $K ={{M_{pl}}^{4}\over{2}}E_{*}^{n}$
and $E_{*}$ measures directly the kinetic contribution of the field.

Inserting into eq.(21) above, and solving for $M(t)$ we obtain, assuming $n > 0$

\begin{equation}
M(t)={M_{i}\over{1-F(t,t_{i})}}
\end{equation}

where
\begin{equation}
F(t,t_{i})={27\pi{{\eta}t_{*}M_{i}}\over{2(n+1)}}\biggl[{(t/t_{*})}^{n+1}-{(t_{i}/t_{*})}^{n+1}\biggr]
\end{equation}

It is easy to show that a set of solutions parametrized by the
initial masses, time constants $t_{*}$ and $n$ exist where a huge
growing of the seed black holes is possible, provided the
constraint given by eq.(19) holds all the time. SMBHs will arise (from initially small PBHs) at a final time $t_{f}$ if 

\begin{equation}
0<1- F(t_{f},t_{i}) \ll \, 1
\end{equation}

Thus, all those PBHs with initial masses of order

\begin{equation}
M_{i}\sim{({t_{f}E_{*}})}^{-(n+1)}\biggl[{2(n+1)E_{*}\over{27\pi{\eta}}}\biggr]
\end{equation}

would end with large masses ($M \sim{10}^{6}M_{\odot}$ or bigger) at
final times $t_{f} \gg t_{i}$.

Numerically, the mass is

\begin{equation}
M_{i} \sim {(n+1)\over{\eta}}
{\biggl[{10^{(24+{2\over{n+1}})}}\biggr]}^{-(n+1)}
{{(t_{f}/s)}^{-(n+1)}}{{(E_{*}/GeV)}^{-n}} \, GeV
\end{equation}

where we have absorbed a coefficient $O(1)$ into the efficiency $\eta$. Initial masses
masses may be large only if the scale $E_{*}$ is extremely small when measured in $GeV$ if $t_{f}$ is inside the radiation-dominated era, according to eq.(29).
 
We may invert the reasoning above and assert that if we had some initial black hole
formed with $M_{i}$ at $t_{i}$, then, the constant $E_{*}$ need to be larger than

\begin{equation}
E_{*}\geq  10^{-{(26+24n)\over{n}}}
\, {\biggl( {t_{f}\over{s}}\biggr)}^{-(1+1/n)}
{\biggl( {(n+1)\over{(\omega_{\eta} M_{i}/GeV)}}\biggr)}^{1/n} 
\, GeV \equiv \, \Theta_{1}
\end{equation}

(with $\omega_{\eta} \equiv (27 \pi \eta/2)$)
for the black hole to grow to the SMBH regime.
Note that when $E_{*}$ is larger, we need smaller initial masses in order
to obtain larger SMBHs at the final time $t >> t_{f}$, as expected.

Eq.(28) also says that our approximations to the actual physical
accretion are valid if and only if

\begin{equation}
M_{i} < {2(n+1)\over{27\pi{\eta}}}{E_{*}{(tE_{*})}^{-(n+1)}}
\end{equation}

and the formulae above stay valid only if the parameter $E_{*}$
does not change with time.
The bottomline of eq.(29) is that if HB+GSL hold for all times, 
then seed PBHs can not
have arbitrary initial masses  (independently of the
details of their formation) if they had to grow by accreting a 
quintessential field within the proposed class. {\it Mutatis mutandis} 
the same conclusions hold for other fuels
for accretion, where we must use eq.(17) instead), in order to
obtain supermassive black holes at the final time $t_{f}$. Note that
this constraint gets weaker with time because
$S_{hp} \propto{t}^{2}$, and this constraint will become 
at some time weaker than the geometric causal condition $r_{g} <
R_{hp}$. The HB is quite restrictive for large masses at black
hole formation, as discussed in Section 3. 
Considerations on the accretion before formation 
must be added to this picture (see next Section).

Finally, according to eq.(19), $E_{*}$ must also satisfy

\begin{equation}
E_{*} \leq 6.6 \times 10^{-2/n}
10^{-(25+24/n)} 
\, {\biggl( {t_{f}\over{s}}\biggr)}^{-(1+1/n)}
{\biggl( {(n+1)\over{( M_{i}/GeV)}}\biggr)}^{1/n}
\, GeV  \equiv \Theta_{2}
\end{equation}

Therefore, only the PBHs contained within the range defined by
$\Theta_{1}(M_{i})< E_{*} < \Theta_{2}(M_{i})$ will satisfy the HB and become SMBHs at late times $t$ simultaneously. 
This leads to the with the following constraint on a positive $n$

\begin{equation}
n < 0.65 \log (27 \pi \eta/2)
\end{equation}

A careful examination of the $n<0$ case leads us to conclude that the lower limit  thus obtained is irrelevant when compared to the $n=-1$ case already  discussed, 
that is, the index is actually limited by $-1$ from below. The case of a constant flux $n=0$ can be also worked out without complications. We conclude that a window of indexes $n$ exist for quintessence to cause the growth of seed PBHs to the SMBH regime. Such a window 
is independent of $M_{i}$. Other physical effects may
be important, for example,  generally speaking, the depletion  of
the quintessence flux around the black hole can not be ignored for
large masses, an effect that has to affect the parameter $\eta$.

%EDITOR PLEASE PUT FIG. 1 HERE

\bigskip
{\bf Causality and Holographic requirements}

As shown in the previous Section, when eqs.(28-30) are satisfied, the energy input by the
accretion of quintessence would be enough to drive a black hole with initial mass $M_{i}$ to values $M_{f} > {10}^{6}M_{\odot}$.

This energy input strongly depends on the initial mass $M_{i}$ and
needs to be very large if the black hole was initially very small.
In addition, if the accretion rate is very high, the
black hole that was initially below the HB will
blow that bound at some point. Then, to keep these black holes
below the HB they must also obey the constraints
$\dot{S}_{smbh}(t)<\dot{S}_{hp}(t)$. However, according to the
eq.(10), this inequality is equivalent to

\begin{equation}
\dot{M}(t) < {{3.4}h_{0}^{2}{M_{pl}}^{4}\over{M(t)}}t
\end{equation}

We solve the evolution from $t_{i}=0$ until $t_{f}>>t_{i}$ as above and then

\begin{equation}
M(t_{f})<M_{i}
{\biggl({{1+{({t_{f}/{\tau}})}^{2}}}\biggr)}^{1/2}
\end{equation}

For $t_{f} \gg \tau = {M_{i}h_{0}^{-1}\over{2.6M_{pl}^{2}}}
\sim {2.5\times{10}^{-6}} s
{({M_{i}\over{M_{\odot}}})}{h_{0}}^{-1}$, eq.(35) becomes $M(t_{f}) <
1.84h_{0}M_{pl}^{2}t_{f}$.

In order to enforce a strictly causal growth $\dot{r_{g}} < c$,
the PBHs must obey also the condition $\dot{M} <
0.5{M_{pl}}^{2}$ at all times. Then, solving for $M(t_{f})$ we obtain

\begin{equation}
M(t_{f})< M_{i}\, 
{\biggl(1+{8}\times{10}^{4}{(t_{f}/s)}{(M_{\odot}/M_{i})}\biggr)}
\end{equation}

In other words, any black hole with initial mass satisfying the
Holographic Bound at $t_{i}$ will eventually be
superholographic at $t=t_{f}$ (that is, $S_{bh}(t_{f})>S_{total}(t_{f})$) 
unless causality holds.

Since that the causality requirement is very strong, we rule out
superholographic black holes in normal circumstances of physical
accretion. If we impose that the solution given by eqs.(25-26) must
satisfy the Holographic Bound for the maximal rate of energy gain,
then combining with eq.(35) 

\begin{equation}
{M_{i} \over{1 - F}} < M_{i}
{\biggl({{1+{({t_{f}/{\tau}})}^{2}}}\biggr)}^{1/2}
\end{equation}

Using eq.(37),  we may obtain an upper bound for the cosmological
time at which our approximations must break down. Defining the
dimensionless quantities $\theta_{1}=(t_{f}/\tau)$,
$\theta_{2}=(t_{f}/\zeta{t_{*}})$ and
$\zeta^{n+1}={2(n+1)/27\pi\eta}$, eq.(37) becomes

\begin{equation}
{\theta_{1}}^{2}-(2{\theta_{1}}^{2}+1){\theta_{2}}^{n+1}+({\theta_{1}}^{2}+1){\theta_{2}}^{2(n+1)}>0
\end{equation}

to be solved for a given $M_{i}$, $\eta$ and $n$ set of values.

We must acknowledge that, in principle, seed black holes may still reach
the SMBH regime without blowing the HB even if the condition 
$\dot{S}_{smbh}(t)<\dot{S}_{hp}(t)$ is not
satisfied, provided its growth effectively stopped while still below the HB value. 
These models, however, must be analyzed in a one-by-one basis to check their viability.

We close this Section with the observation that both the HB
requirement and the relativistic bound $\dot{r_{g}}<c$  on
$\dot{M}$ lead to essentially the same value $\dot{M} < 2
\times{10}^{38}g{s}^{-1}$ within a numerical factor of the order
of one.

\bigskip
{\bf Conclusions}

We have discussed a possible form to limit the formation times of
primordial black holes formed directly or grown by accretion which
may be residing at the center of most galaxies as recently
identified by a variety of observations. The huge entropy
contained in these SMBH allows to limit their formation quite
efficiently, since the total content of entropy of the universe is
likely to be bounded by the HB. Even if preliminar,our analysis
of the quintessence models for the growth of seed PBHs has been
found to leave room for their formation and further growth,
although not for arbitrary fluxes. The general argument developed
in Ref. [12] against fast-growing solutions for PBH growth with
radiation flux $\varrho_{rad}(t)\propto{t}^{-2}$ can be directly
applied to the the particular model  involving quintessence flux
$\dot{\varphi}^{2}\propto{t}^{-2}$. Generally speaking, the
quintessence model must allow the flux to decrease slower than
$t^{-2}$ for PBHs to grow at all, and to stay constant or increase
for substantial accretion to occur, as needed for achieving the
SMBH condition in a short time. General conditions on $\dot{M}$
have been obtained by a combination of causal and holographic
arguments and are amenable of specific applications.

Other models can be constructed to  produce a population of
SMBH starting from seed PBHs. For example, accretion in a
brane-world high-energy phase has been recently studied [10] and
shown to allow a substantial growth in which $\dot{M} \propto \,
M/t$. It may be possible to arrive to the end of the high-energy
phase with very massive black holes, although the full
consequences of this scenarios are yet to be explored.

\bigskip
{\bf Acknowledgments}

Both authors wish to thank the S\~ao Paulo State Agency FAPESP for financial
support through grants and fellowships. J.E.H. has been partially supported by 
CNPq (Brazil). An anonymous referee is acknowledge for criticisms that 
helped to improved the original draft.

\bigskip
{\bf References}

[1] B.M.Peterson, {\it An introduction to Active Galactic Nuclei}, Cambridge, 1997.

[2] F. De Paolis, G. Ingrosso, A.A. Nucita, D. Orlando, S. Capozziello and G. Iovane,
{\it Astron. Astrophys.}{\bf 376}, 853 (2002).

[3] T.Kephart and Y.Jack Ng, gr-qc/0204081 (2002).

[4] L. Ferrarese, {\it Ap.J.}{\bf 578}, 90 (2002) ; J. Silk, {\it Space Sci. Rev.}{\bf 100}, 41 (2002).

[5] D.Bigatti and L.Susskind, hep-th/0002044 (2000) and N.Ohta, gr-qc/0205036 (2002).

[6] J.Bekenstein, {\it Phys.Rev.D}{\bf 23}, 287 (1981).

[7] E.Verlinde, hep-th/0008140 (2000).

[8] J.Bekenstein, {\it Phys.Rev.D}{\bf 9}, 3292 (1974).

[9] R. Bean and J.Magueijo, astro-ph/0204486 (2002).

[10] R. Guedens, D. Clancy and A.R. Liddle, astro-ph/0208299

[11]  J.D.Barrow, E.J. Copeland and A.R. Liddle, {\it Phys. Rev. D}{\bf 46}, 645 (1992);
see also {\it MNRAS} {\bf 253}, 675 (1991).

[12] P.Custodio and J.E. Horvath, gr-qc/0203031 (2002).

[13] P.Custodio and J.E. Horvath, {\it Phys.Rev.D}{\bf 58}, 023504 (1998).

\vfill\eject

\noindent
Figure captions

\bigskip
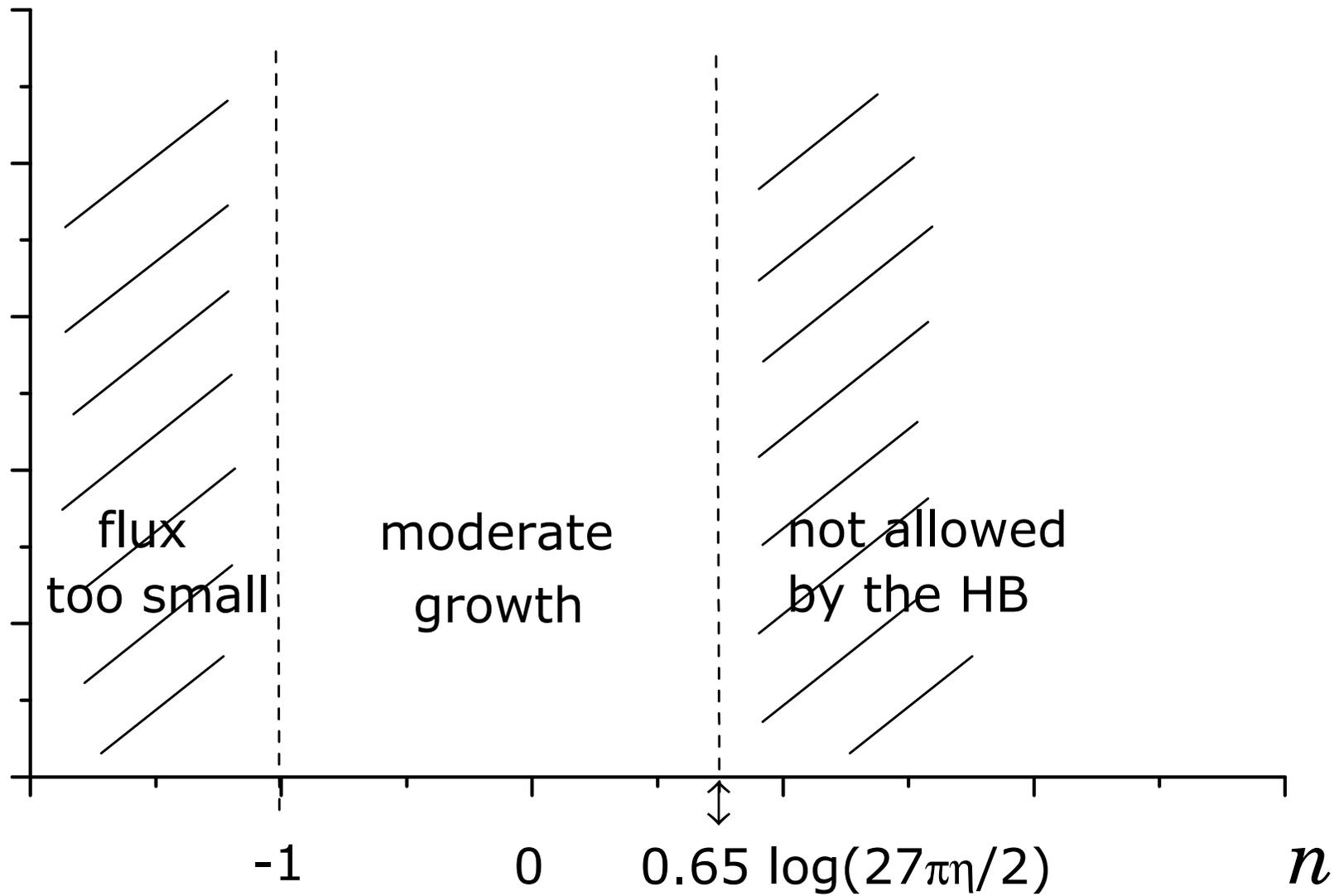
Fig. 1. The window of indexes $n$ that allow a growth of small PBHs to the SMBH regime.
As explained in the text, the upper bound is set by the HB requirement on $E_{*}$ (eq.32); while the lower bound $n=-1$ is imposed by the requirement of having enough flux of fuel to complete the process (Ref.12). The possible modes of growth include the constant quintessence flux case $n=0$, detailed conditions to be satisfied by these particular power-law models are given in the text.

\end{document}